   \documentclass[prl,aps,twocolumn,showpacs]{revtex4}
\usepackage[dvips]{graphicx}
\usepackage{dcolumn}
\newcommand{\be}{\begin{equation}}
\newcommand{\ee}{\end{equation}}
\newcommand{\bea}{\begin{eqnarray}}
\newcommand{\eea}{\end{eqnarray}}
\newcommand{\nn}{ \nonumber}

\begin{document}
\topmargin=-15mm

\title{The Effect of a Magnetic Field on the Acoustoelectric current in a Narrow Channel} 

\author{Natalya A. Zimbovskaya* and Godfrey Gumbs}

\affiliation{ Department of Physics and Astronomy, Hunter College, City University of New York, 695 Park Avenue, New York, NY 10021}

\begin{abstract}
The effect of a perpendicular magnetic field on the quantized current induced
by a surface acoustic wave in a quasi-1D channel is studied. The channel has been produced experimentally in a GaAs heterostructure by shallow etching techniques and by the application of a negative gate voltage to Schottky split gates. Commensurability oscillations of the quantized current in this constriction have been observed in the interval of current between quantized plateaus. The results can be understood in terms of a moving quantum dot with the electron in the dot tunneling into the adjacent two-dimensional region. The goal is to explain qualitatively the mechanism for the steplike nature of the acoustoelectric current as a function of gate voltage
and the oscillations when a magnetic field is applied. A transfer Hamiltonian formalism is employed.  
    \end{abstract}

\maketitle

\section{1. INTRODUCTION}

In recent experiments, it has been shown that when a surface acoustic wave (SAW) is launched on a piezoelectric
heterostructure, such as GaAs/AlGaAs, containing a two-dimensional electron gas (2DEG) with a narrow channel, a quantized acoustoelectric current may be measured \cite{one,two,three}. The narrow (quasi-one-dimensional) channel may be formed either by shallow etching techniques or by the application of a negative gate voltage to Schottky split gates. The acoustoelectric current through the narrow
channel has been observed beyond pinch-off. In the
experiments of Refs. \cite{one,two,three}, it was seen that the acoustoelectric current exhibits steplike behavior as a function of the applied gate voltage within a certain frequency range of the SAW. Furthermore, the current on a plateau is given by $ I = efn $, with an accuracy of one part per $10^4$, where $n = 1,2,3 \dots, $ and where $e $ is the electron charge, and $ f $ is the SAW frequency. The quantized acoustoelectric current can be explained by an electron being scooped up from the source region and getting trapped in a moving quantum dot formed by the electric potential of the SAW and the potential barrier within the channel. The electrons are transferred by the SAW and the potential barrier within the channel. The electrons are transferred by the SAW through the channel over the intrachannel barrier. A detailed theory of the experiments reported in references \cite{one,two,three} was provided by Aizin {\it et al} \cite{four,five}, when no external magnetic field was applied.

More recent experiments by Cunningham {\it et al} \cite{six} were done to determine the effect of a perpendicular magnetic field on the quantized acoustoelectric current in GaAs/AlGaAs heterostructures. It was shown in Ref. \cite{six} that at a fixed gate voltage the acoustoelectric current oscillates as a function of magnetic field for $ B \stackrel{<}{\sim} 0.2$T.
The amplitudes of these oscillations are larger for values of the gate voltage between the current plateaus compared
with their values near the plateaus. It was suggested in \cite{six} that these oscillations were commensurability oscillations due to the geometrical resonances for the cyclotron orbits of the 2D electrons outside the channel with the SAW wavelength $ \lambda$. That is, the nature of the oscillations for the acoustoelectric current is the same as that of that of the well-known Weiss oscillations which were observed for the magnetoresistivity of a modulated 2DEG \cite{seven,eight}. Commensurability oscillations of a non-quantized acoustoelectric current were reported in \cite{nine}. Admittedly, a systematic theoretical study of the geometrical oscillations of the quantized acoustoelectric current is a very complicated problem and cannot be solved analytically. Numerical solution also has its challenges. In this work, our goal is to use a very simple model which enables us to give a semiquantitative analysis of the magnetic field effect. Our results are in qualitative agreement with the experimental results in \cite{six}.

\section
{2. The model and numerical results }

In our formulation of the problem, we choose coordinate axes such that the 2DEG is in the $x y $ plane and the quasi-one-dimensional channel lies along the $x$-axis. The electrostatic potential induced by the gate is $ U(x) $ with an effective barrier height above the Fermi energy of the 2DEG taken as                 
   \begin{equation} 
U_0 = \frac{\hbar^2}{2m l^2_0}\ \label{e1}
    \end{equation}
  where $ m $ is the electron effective mass and the parameter $l_0$ characterizes the height of the electrostatic potential barrier. The SAW is also launched along the $x$ axis and the accompanying electric potential is taken simply as 
   \begin{equation} 
\Phi (x,t) = \Phi_0 \cos (kx - \omega t)\ \label{e2}
     \end{equation}
   where $ k = 2\pi/\lambda $ is the wave vector, $\omega = 2\pi f$ is the angular frequency of the SAW, and $ \Phi_0 $ is the reduced magnitude of the SAW-induced potential which is strongly screened due to a high conductivity of the 2DEG.

In the channel, the electric potential of the SAW is modified from its simple harmonic form (2) owing to screening by the metal gates as well as the presence of the intrachannel potential barrier $ U (x)$ \cite{four}. As a result, a one-dimensional quantum dot is formed at the entrance of the channel. The size of the dot along the $x$-direction is less than the SAW wavelength and this confining potential quantizes the captured electron energy levels. Both depth and shape of the quantum dot moving through the channel are time dependent. Furthermore, we assume that the quantum dot moves slowly enough that a captured electron can adjust to a time-varying electric potential. In this adiabatic approximation, we can treat time as a parameter in our calculations.

We first consider the case where there is no magnetic field present, to show that the simplified consideration presented here gives results which are in a qualitative agreement with the experiments of \cite{one,two,three} and with the detailed theoretical analysis of \cite{four,five}. In the absence of the external magnetic field, the moving quantum dot can scoop up electrons from the surface of the Fermi sea outside the channel. Here, we restrict our attention to the case where only one elecron from the Fermi surface is
captured by each quantum dot formed by the SAW-induced electric potential near the intrachannel barrier. The acoustoelectric current is then given by
   \begin{equation} 
I= (1 - P(0))e f \label{e3}
       \end{equation} where $ P(0) $ is the tunneling probability for the captured electron to return to the source during one SAW cycle in the absence of the external magnetic field. When the well-known tunneling probability \cite{three,four} is averaged over the SAW period we obtain
  \bea 
&& P(0) = \frac{1}{2 \pi} \int_0^{2\pi} d \theta \\  \nn \\ &\times& \exp \left (- \frac{1}{\hbar}
\int_{X_1 (\theta)}^{X_2(\theta)}
\sqrt {2m (V_{tot}(x,\theta)- \epsilon)}dx \right ) \nn 
 \label{e4}
      \eea
  where $\theta=\omega t, \; X_1 (\theta) $ and $ X_2 (\theta) $ are coordinates for the forbidden region at time $t$ when the electron is in the lowest state with energy $\epsilon$ in the dot, and $V_{tot}(x, \theta)$ is the total potential energy determining the electron eigenstates in the well.


  As the quantum dot moves up to the top of the potential barrier within the channel $ (0 < \theta < \pi), $ the probability for the captured electron to tunnel back to the source is, in general, much larger than the probability for it to tunnel to the drain through the barrier. For a pinched-off channel the latter is negligible. However, as the dot moves down from the top of the barrier $ ( \pi < \theta < 2 \pi), $ the probability for tunneling back to the source is much less than the probability of tunneling to the drain. This asymmetry mainly arises due to the difference in the tunneling path lengths arising from the combination of the electric potential of the SAW and the potential barrier within the channel. Except for a short time interval when the quntum dot is near the top of the potential barrier within the channel, the tunneling path between the dot and the drain (source) is significantly longer for $ 0 < \theta < \pi \ (\pi < \theta < 2 \pi ) $ than the tunneling path from the dot to the source (drain). The narrower forbidden region is mainly formed by the SAW-induced potential. Consequently, the dominant contribution to the tunneling probability $ P(0) $ arises from the first half of the SAW cycle $(0<\theta <\pi )$. A detailed numerical analysis in \cite{four} verifies these results. Therefore, we time-average the expression for $ P(0)$ within the interval $ 0 < \theta < \pi,$ neglecting a small correction which originates from the remaining part of the SAW period. Within this time interval, $ P(0)$ is mostly determined by the tunnelling of the electron through the barrier formed by the SAW-induced potential.

To proceed in obtaining a simple qualitative estimate of the acoustoelectric current, we approximate the ground-state energy of the captured electron by the well minimum. Furthermore, we simulate the SAW-induced potential within the channel by a simple harmonic form with the shorter wavelength and a smaller amplitude. This corresponds to a smaller region for forming the quantum dot and a higher position of the ground state of the electron captured there. A higher intrachannel barrier produces a stronger distortion of the original SAW-induced potential. To take this into account, we introduce the following simple model of the SAW-induced potential in the channel:
     \begin{equation} 
\Phi = \Phi_{0}^* \cos (k^*x - \omega t)
\label{e5}
\end{equation}
  where
  \[ 
   \Phi_0^* = \Phi_0 \left (\frac{\beta}{\beta + 1} 
\right )^\alpha \quad k^* = k \left( \frac{\beta + 1}{\beta} \right )^\nu  
  \quad   \alpha, \nu > 0
    \]
  and the dimensionless parameter $ \beta $ is the ratio of the screened SAW-induced potential to the effective height of the intrachannel berrier. When the gate-induced electrostatic potential is weak, $ \beta \to \infty $ and we have $ \Phi_0^* \to \Phi_0 $ and $ k^* \to k. $ Therefore, in this limit when there is no channel, our model (5) correctly describes the SAW. On the other hand, when the barrier is high, $ \beta $ tends to zero. In this case, both the effective height of the SAW-induced barrier $ \Phi_0^* $ and the length of the tunneling path below it, $ 2 \pi/k^*, $ also tend to zero. As a result, $ P(0) \to 1, $ and there is no acoustoelectric current, as we show below (see equation (7)). These conclusions agree with the experimental data of \cite{one,two,three} as well as detailed theoretical and numerical calculations carried out in \cite{four}. Here, we use the model (5) instead of the model used in \cite{four} because it allows us to explain the measurements for the acoustoelectric current in a uniform perpendicular magnetic field. Reasonable agreement between the results presented in this paper and those of \cite{four} can be obtained with a suitable choice of the values of $ \alpha $ and $ \nu. $

Within our model we put the electron at the bottom of the dot. Therefore, the difference $ V_{tot} (x, \theta) - \epsilon $ in the integrand of (4) does not explicitly depend on the gate-induced intrachannel potential and equals $ \Phi_0^* (\cos (k^* x - \theta ) + 1) .$ A straightforward calculation gives
      \bea 
&& P(0) = \frac{1}{\pi} \int_0^{\pi} d \theta \\ \nn \\
& \times& \exp \left [- \frac{\sqrt \beta}{k^* l_0} \left (\frac{\beta}{\beta + 1} \right)^{\alpha/2}
\int_{\pi + \theta}^{3\pi + \theta} \sqrt {\cos (u - \theta) + 1} d u \right ] \nn
\label{e6}
           \eea 
  where $ u = k^*x $. Carrying out the integration in (6), we obtain
     \begin{equation} 
P(0)=\exp \left[- \frac{4\sqrt{2}}{k l_0}\frac{\beta^{\gamma +
1/2}}{(\beta+1)^\gamma}\right]=\exp \bigg(\frac{-\mu}{k l_0} \bigg) .
            \end{equation}  
Here $ \gamma = \nu + \alpha/2. $

In Fig.1, we use (7) to plot $ I/ef=1-P(0)$ as a function of $ \beta $ for $ \lambda = 3 \mu m ,\ \gamma = 3.5, \ l_0 =0.1\mu m$. Our results agree qualitatively with \cite{one,two,three,four} and have a step. Therefore, our simple model calculation produces the basic result that the acoustoelectric current is quantized as a function of the gate voltage or
SAW-induced potential. We now extend this approach to describe the oscillations of the acoustoelectric current in the presence of an external magnetic field directed perpendicularly to the plane of the 2DEG.


In the presence of a magnetic field $ B $, we use the adiabatic approximation and treat the SAW-induced electric potential as a quasistatic electric modulation of the 2DEG outside the channel. The Hamiltonian is \cite{eight}
  \begin{equation} 
H= - \frac{\hbar^2}{2 m} \frac{d^2}{d x^2} + \frac{m}{2} \omega_c^2 (x -
x_0)^2 + \Phi_0 \cos (kx - \omega t) \label{e8}
       \end{equation} where $m $ is the electron effective mass, $\omega_c = e {\bf B}/m $ is the cyclotron frequency and $x_0 = k_y l^2 $ is the
guiding center of the cyclotron orbit, with $ k_y $ the transverse wavenumber and $ l = \sqrt{\hbar/e {\bf B}} $ the magnetic length. The electron energy levels are obtained within the adiabatic approximation by solving the stationary Schr\"odinger equation for the Hamiltonian (8) and treating the time $ t $ as a parameter. The energy eigenvalues in first-order perturbation theory for the electric modulation induced by the SAW are well-known and given by \cite{seven,eight,ten,eleven}
    \be 
E_{n, x_0} = \hbar \omega_c \left (n + \frac{1}{2} \right ) + \delta E_{n,x_0} \label{e9}
  \ee     where
         \begin{equation} 
\delta E_{n, x_0} = \Phi_0 \cos (k x_0 - \omega t) e^{-W/2} L_n (W) .
\label{e10}
        \end{equation}  In this notation, $n$ is a quantum number labelling the Landau levels, $ W = l^2 k^2/2 , $ and $ L_n (W) $ is a
Laguerre polynomial. The modulation lifts the degeneracy and
broadens each Landau level into a band of width $ 2 \Phi_0 e^{-W/2} |L_n (W)|$. 
  \begin{figure}[t]
\begin{center}
\includegraphics[width=7cm,height=7cm]{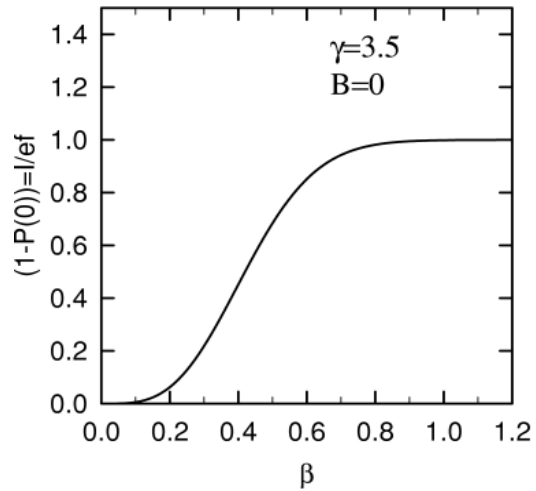}
\caption{ A plot of $ I /e f = 1 - P $ as a function of $\beta $ for $ \lambda = 3 \mu m, \ l_0 = 0.1 \mu m, $ and $ \gamma = 3.5. $ } \label{rateI}
\end{center}
\end{figure}

Let us first concentrate on the first term of the equation (9). In the experiments of reference \cite{six}, we have $ n_e \sim 10^{12}$ cm$^{-2}; \; B \sim 0.1$ T; $ m \approx 0.067 m_0 $ where $ n_e $ is the electron density, and $ m_0 $ is a free electron mass, so the Fermi energy $ E_F $ is of the order of $ 10 me V, $ and $ \hbar \omega_c << E_F.$ When we vary the magnetic field, the number of Landau levels below the Fermi surface changes and the energy of the highest Landau level oscillates as a function of the magnetic field. This gives rise to quantum oscillations of various observables, e.g. Shubnikov-de-Haaz oscillations of the 2DEG magnetoresistivity. The amplitude of quantum oscillations strongly depends on temperature. For moderately low temperatures when $ \hbar \omega_c < 2 \pi^2 k_B T \; (k_B $ is the Boltzmann's constant) the quantum oscillations have a simple harmonic form and small amplitude. The experiments of \cite{six} were carried out at $ T = 1.2$ K. At magnetic fields of the order of 0.1 T, this gives us $ \hbar \omega_c/2 \pi^2 k_B T \sim 0.1.$
Therefore, in our calculations we may assume that the quantum
oscillations are smeared out, and we can identify the energy of the highest Landau level with the Fermi energy $ E_F $ in the absence of the external magnetic field. The second term in the expression (9) gives rise to the well-known Weiss oscillations in the magnetoresistance of a modulated 2DEG \cite{seven,eight,ten,eleven,twelve}. The amplitude of these commensurability oscillations also depends on temperature. However, it has been shown \cite{twelve} that Weiss oscillations can be observed at higher temperatures when $2 \pi^2 k_B T < (k_F/k) \hbar \omega_c ,\ (k_F $ is the Fermi wavenumber). For the experiments \cite{six} we obtain that the commensurability oscillations may be observed when $ T < 6 K,$ which is significantly larger that the actual temperature, and we can neglect the effect of temperature smearing of these oscillations.

To proceed we note that in the absence of a magnetic field, the electrons most likely to be scooped up by the SAW are those which move in the direction in which the SAW is launched and have velocity $v_x $ close to the SAW speed $ s $ \cite{thirteen}. In the experiments \cite{one,two,three,six}, the Fermi velocity of 2D electrons $ v_F $ is much larger than $s. $ Consequently, the probability is high for a captured electron to have its transverse component of velocity $ v_y $ close to $ v_F $. Such electrons move nearly transverse with respect to the channel. After being captured in the quantum dot, they are dragged through the channel at speed $ s. $ At the same time, their transverse motion is confined by the walls of the channel with multiple reflections.

In the presence of a weak perpendicular magnetic field with $ R > \lambda \ (R = v_F /\omega_c $ is the cyclotron radius), electrons cannot be captured by the SAW away from the channel due to their motion along cyclotron orbits which are large compared to the SAW wavelength. However, electrons moving nearly at right angles near the entrance to the channel $(v_x \approx s, \ v_y \approx v_F) $ can be scooped up there and dragged inside as in the case when there is no magnetic field. The electrons are multiply reflected from the walls of the channel and paths within the cannel consist of arcs of their cyclotron orbits. We consider a very narrow channel whose width $ d $ along the $ y$-direction is much less than its length. In the experiments of \cite{one,two,three,six}, the length of the channel is of order of $ \lambda ,$ and in a weak magnetic field $ B < 0.2 $ T, $ R > \lambda, $ and we have $ R >> d. $ A simple geometrical consideration enables us to conclude that under the condition $ d^2 /R << 2 \pi /k^*, $ an electron in the channel can be localized in the moving quantum dot. This means that we can disregard the coupling of the $x$- and $y$-motion as an electron moves through a very narrow channel in the presence of a perpendicular magnetic field.

We choose the entrance to the channel at $ x = 0. $ Then the centre of the cyclotron orbit for an electron which moves nearly transversely at the entrance to the channel is located at a distance close to $ R $ from the entrance. As a first approximation, we can put $ x_0 = - R $ in expression (10) for the correction to the electron energy arising from the SAW-induced electric potential. As described above, electrons from the highest occupied energy level can be scooped up at the entrance of the channel. However, in the presence of a magnetic field we cannot treat the corresponding energy as being a constant because it contains the oscillating term given in (10). Therefore, the effective intrachannel barrier height above the highest electron energy before the barrier now depends on time as well as the magnetic field. Thus we can write the following expression for the effective height of the intrachannel barrier at the presence of the magnetic field:
   \begin{equation} 
U= U_0 (1 - \beta \cos (kR + \omega t) e^{-W/2} L_n (W)) .
  \label{e11}
\end{equation} Correspondingly, the parameter $ \beta $ in equation (6) has to be replaced by
    \begin{equation} 
\beta^* = \frac{\beta}{1 - \beta \cos(kR + \omega t) e^{-W/2} L_n (W)}
           . \label{e12}
\end{equation} Within the framework of our semiquantitative approach, we have reduced our problem to a one-dimensional one, and we can calculate the tunneling probability in the presence of the magnetic field using the expression (6) where $ \beta $ is replaced by $ \beta^* $ in the expressions for parameter $ k^* $ and $ \Phi_0^*. $ Expanding the exponent of the integrand over $ \theta $ in (6) in powers of a small
parameter $ [ \beta/(1 + \beta)] \cos (k R +\omega t) e^{-W/2} L_n (W)$ and keeping two first terms of the expansion we obtain after a straightforward calculation
   \begin{equation} 
P(B) = P (0) \frac{1}{\pi} \int_0^\pi d \theta
\exp (-u \cos(kR + \theta ) )
 \label{e13}
          \end{equation} where $ u = (\mu \rho /k l_0) e^{-W/2} L_n (W) $ and $ \rho \equiv \beta \gamma/(1 + \beta). $ When $ u << 1 $, we have
  \begin{equation} 
P(B) \approx P(0) \left [1 + \frac{2 \mu \rho}{\pi k l_0} e^{-W/2} L_n (W) \sin (kR) \right ].
           \label{e15}
       \end{equation} We consider here moderately weak magnetic fields for which $ E_F /\hbar \omega_c >> 1$ and $ k R > 1 $. Therefore, we can use the asymptotic approximation for the Laguerre polynomial to simplify our expression for the tunneling probability. As a result we obtain:      \begin{equation} 
P(B) \approx P(0) \left [1 + \frac{2 \mu \rho}{\pi k l_0}
J_0 (kR) \sin (kR) \right ] .
        \label{e16}
       \end{equation} It follows from (15) that the quantized acoustoelectric current
can have an oscillatory dependence on the applied magnetic
field. When $kR\gg 1$, the oscillations coincide in
period and phase with the Weiss oscillations which were observed in a 2DEG modulated by a weak electrostatic potential.

The oscillations can be noticeable for moderately small values of the parameter $ \beta $ providing intermediate values of probability $ P(0) $ which are close neither to zero nor to unity. For such values of $\beta $ the acoustoelectric current takes non-zero values considerably smaller than $ef$ on the first plateau. This is in qualitative agreement with the experiments of \cite{six}. In figure 2, we plot the acoustoelectric current as a function of magnetic field for $ \beta = 0.3,0.5,0.9, $ and 1.5. In our calculations, we used (15) for the tunneling probability $P$ and chose
the SAW wavelength, Fermi energy, the effective height of the barrier within the channel, and the electron effective mass about the same as those reported in \cite{six}. We used $ \lambda = 3 \mu m , \ E_F = 10\ mV, \ l_0 = 0.1 \mu m , \ m = 0.067m_0 $. Our results in Fig.2 are also in qualitative agreement with the experimental results of \cite{six}.

\begin{figure}[t]
\begin{center}
\includegraphics[width=7.2cm,height=7cm]{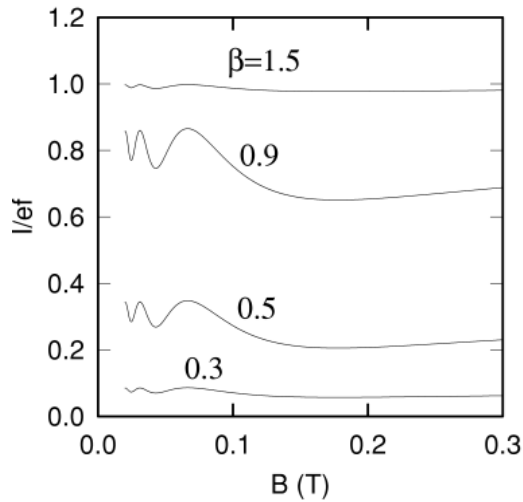}
\caption{A plot of $ I /e f = 1 - P $ as a function of magnetic field $ B $ for $ \lambda = 3 \mu m, \ E_F = 10 mV, \ l_0 = 0.1 \mu m, \ m = 0.067 m_0, \ \beta = 0.3,0.5,0.9, $ and 1.5, and $ \gamma = 3.5. $
} \label{rateI}
\end{center}
\end{figure}

\section{ III. SUMMARY AND CONCLUSIONS}

\vspace{0.25in}

We have presented the results of our calculations simulating
recent experimental data clearly showing commensurability oscillations of the quantized acoustoelectric current in
the presence of an external magnetic field. The oscillations were discovered in measurements of the acoustoelectric current for which a quantum dot captures an electron and transports it through a quasi-one- dimensional pinched-off channel in GaAs/AlGaAs heterostructures. In this letter, we presented a simple semiquantitative model which allows us to describe the effect of the quantization of the acoustoelectric current in the absence of a magnetic field as well as the geometric oscillations in the presence of a magnetic field. We have shown that these oscillations have the same nature as Weiss oscillations for magnetotransport in a modulated 2DEG. In our calculations described here, a weak electrostatic modulation is created by the electric field accompanying the SAW. Our model provides qualitative agreement with the results of
experiments \cite{one,two,three,six}, thereby confirming that we have included the essential features of the effect described.
\vspace{2mm}

{\bf Acknowledgments:}

The authors gratefully acknowledges the support in part
from the City University of New York PSC-CUNY-BHE grants \#666414, \#664279 and \#669456 as well as grant \# 4137308-01 from the NIH. NAZ also thanks G.M. Zimbovsky for help in preparing the manuscript.


*Present address: Department of Physics and Electronics, University of Puerto Rico - Humamao.

\end{document}